\documentclass[amsmath,amssymb,aps,prd,twocolumn,superscriptaddress,nofootinbib,showpacs]{revtex4-1}
\usepackage{graphicx}
\usepackage{dcolumn}
\usepackage{bm}
\usepackage{amsmath}
\usepackage{amssymb}
\usepackage{xcolor}
\usepackage{bbm}
\usepackage{hyperref}
\usepackage{booktabs}
\usepackage{tabularx}
\usepackage[utf8]{inputenc}
\usepackage[T1]{fontenc}
\begin{document}

\title{Scalar--Tensor Gravity as a Probe of Generalized Black Hole Entropy}
\author{Hussain Gohar}
\email{hussain.gohar@usz.edu.pl}
\affiliation{Institute of Physics, University of Szczecin, Szczecin, Poland}

\date{\today}

\begin{abstract}
We develop a geometric realization of a broad class of generalized black hole entropy functionals by establishing a systematic correspondence with the Misner–Sharp quasilocal mass and the Wald Noether-charge entropy in scalar–tensor theories of gravity. The resulting models feature a scale-dependent effective gravitational coupling, whose functional dependence is determined by the underlying entropy parameters. Within this framework, we derive explicit Einstein-frame scalar potentials: for Barrow entropy, a steep exponential potential; for Tsallis–Cirto entropy, an exponential potential governed by the nonextensivity parameter; and for quantum-gravity and entanglement-induced corrections, an approximately linear potential. These distinct potentials generate characteristic cosmological phenomenology, with implications for inflationary dynamics, late-time dark-energy behavior, and non-singular bouncing cosmologies. The framework is compatible with current constraints from solar-system tests, big-bang nucleosynthesis, and pulsar-timing observations, and it yields predictions that can be probed by forthcoming observational surveys. In this way, the analysis establishes a unified and geometrically grounded connection between information-theoretic entropy proposals and gravitational field theory.
\end{abstract}

\maketitle

\section{Introduction}

The Bekenstein--Hawking entropy relation~\cite{Bekenstein:1973ur, Hawking:1974rv},
$S_{\text{BH}} = \frac{k_B c^3 A}{4\hbar G}$,
asserts that the entropy of a black hole is proportional to the area $A$ of its event horizon, where $k_B$, $c$, $G$, and $\hbar$ denote Boltzmann’s constant, the speed of light, Newton’s gravitational constant, and the reduced Planck constant, respectively. Rooted in Hawking’s area theorem~\cite{Hawking:1971vc} and further supported by recent gravitational-wave measurements~\cite{KAGRA:2025oiz, Tang:2025jyj}, this law provides the cornerstone for interpreting gravitational dynamics in thermodynamic terms. In particular, Jacobson derived Einstein’s field equations from the Clausius relation~\cite{Jacobson:1995ab}, Verlinde proposed an emergent description of gravity as an entropic force~\cite{Verlinde:2010hp}, Padmanabhan obtained the Friedmann equations from a principle of holographic equipartition~\cite{Padmanabhan:2003pk}, and Cai and Kim rederived the Friedmann equations from the first law of thermodynamics applied to cosmological horizons~\cite{Cai:2005ra}.

The Bekenstein--Hawking area law is, however, widely regarded as only the leading-order term in the entropy of a black hole. Quantum-gravitational corrections, entanglement entropy, fractal or quantum-corrected horizon geometries, and nonextensive generalizations of statistical mechanics all motivate modified entropy functionals~\cite{Rovelli:1996dv, Meissner:2004ju, Medved:2004yu, Das:2007mj, Adler:2001vs, Alonso-Serrano:2018ycq, Barrow:2020tzx, Tsallis:1987eu, Tsallis:2012js, reny1, Czinner:2015eyk, Alonso-Serrano:2020hpb, Kaniadakis:2005zk, Nojiri:2022aof, Nojiri:2022dkr}, which have been extensively employed in holographic contexts and within the broader gravity–thermodynamics correspondence~\cite{Li:2004rb, Komatsu:2013qia}. In particular, it has recently been shown that adopting a generalized entropy within the gravity–thermodynamics correspondence generically induces horizon-area–dependent modifications of the field equations and an effectively varying gravitational coupling~\cite{Lu:2024ppa, DiGennaro:2022grw}; the present work develops this connection systematically at the level of the mass-to-horizon relation, the Misner–Sharp quasilocal mass, and the Wald entropy, and derives the resulting Einstein-frame scalar dynamics. These generalized entropies, in turn, raise several conceptual issues~\cite{Nojiri:2021czz, Cimdiker:2022ics, Gohar:2023hnb, Gohar:2023lta, Gohar:2025yfx}: What is the appropriate notion of temperature to associate with a given generalized entropy functional? Is it possible to maintain holographic thermodynamic consistency while retaining the Hawking temperature, derived from quantum field theory in curved spacetime? And for each entropy modification, what underlying geometric or gravitational theory yields a consistent classical realization?

The first two of these questions are directly tied to the requirement of holographic thermodynamic consistency. Within any framework implementing the holographic principle~\cite{tHooft:1993dmi, Susskind:1994vu}, the entropy $S$ and temperature $T$ assigned to a holographic horizon must, when inserted into the Clausius relation
$dE = T\,dS$, ensure compatibility between the energy $E$ and the mass $M$ attributed to the holographic horizon. This consistency condition is generically violated when generalized entropy functionals are combined in a naive manner with the standard Hawking temperature $T_h$ in the Clausius relation~\cite{Cimdiker:2022ics, Nojiri:2021czz}.

One proposed resolution is to promote $T_h$ itself to a generalized, entropy-dependent temperature, constructed so as to formally restore holographic thermodynamic consistency for each specific entropy prescription~\cite{Cimidiker:2023kle}. This strategy, however, is not supported by quantum field theory in curved spacetime: among such candidates, only the usual Hawking temperature admits a rigorous derivation from quantum field theory via the surface gravity $\kappa$ defined on the relevant horizon. Replacing $T_h$ by an ad hoc modified temperature therefore abandons precisely the quantum-field-theoretic underpinnings that render the Clausius relation a physically well-motivated bridge between gravity and thermodynamics.

The rest of the paper is organized as follows. In Sec.~\ref{sec:mhr} we introduce the generalized mass-to-horizon relation, derive the associated entropy functional, and establish the positivity and branch-selection criteria, including the emergence of a minimum horizon radius and its interpretation in terms of black hole remnants. Section~\ref{sec:geometric} develops the geometric realization of the framework in scalar--tensor gravity via the Misner--Sharp quasilocal mass, derives the corresponding Jordan- and Einstein-frame scalar potentials, and verifies the consistency of the construction with the Wald entropy. In Sec.~\ref{sec:jacobson} we provide an independent thermodynamic justification of the framework, based on Jacobson's derivation of the gravitational field equations as an equation of state. Section~\ref{sec:varyG} confronts the resulting scale-dependent gravitational coupling with solar-system, big bang nucleosynthesis, and pulsar-timing constraints, while Sec.~\ref{sec:cosmo} explores the cosmological implications of the entropy-induced scalar potentials for inflation, dark energy, and non-singular bouncing scenarios. Finally, Sec.~\ref{sec:summary} summarizes our results and outlines directions for future work.

\section{The generalized mass-to-horizon relation}
\label{sec:mhr}

To address these issues, in~\cite{Gohar:2025yfx, Gohar:2023lta, Gohar:2023hnb}
we introduced the notion of a fundamental \textit{mass-to-horizon relation} (MHR),
which provides a systematic route for constructing generalized entropy functionals
that remain fully compatible with the standard Hawking or Cai-Kim\footnote{The Hawking-like temperature associated with cosmological horizons is justified within quantum field theory; see, for example, Refs.~\cite{Cai:2005ra,Cai:2008gw,Li:2008gf}.} temperature.
In particular, we proposed the generalized MHR~\cite{Gohar:2025yfx}
\begin{equation}
M = \gamma \, \frac{c^2}{G}\,\ell_{\rm Pl}
\left[ \frac{L}{\ell_{\rm Pl}} \;\mp\; \beta
\left(\frac{L}{\ell_{\rm Pl}}\right)^{3-\alpha} \right]^m,
\label{eq:MHR}
\end{equation}
where $\gamma > 0$ and $\beta > 0$ are dimensionless constants, $m > 0$ and
$\alpha \in \mathbb{R}$ are dimensionless exponents, and
$\ell_{\text{Pl}} = \sqrt{\hbar G / c^3}$ denotes the Planck length. The upper (lower) sign
corresponds to a positive (negative) subleading deviation from the standard linear
scaling $M \propto L$ \cite{Gohar:2023hnb}; physically, these deviations encode quantum corrections that
decrease (increase) the effective mass enclosed by the horizon with respect to its
standard value. Within this generalized MHR framework, the Clausius relation guarantees
that $T_h$ and the associated generalized entropy $S_{\rm BH}$ are thermodynamically
consistent.

Implementing the Clausius relation with $T_h = \hbar c / (2\pi k_B L)$ and expanding to
leading order in $\beta$, the generalized MHR gives rise to the entropy functional
\begin{equation}
S_G = 2\pi k_B \gamma \left[\frac{m}{m+1}
\left( \frac{L}{\ell_{\rm Pl}} \right)^{m+1}
\mp\, \frac{m(\sigma - 1)}{\sigma}\,\beta
\left( \frac{L}{\ell_{\rm Pl}} \right)^{\sigma}
\right],
\label{eq:SG}
\end{equation}
where $\sigma \equiv m + 3 - \alpha$. In the limiting case $\sigma \to 0$ (i.e.,
$\alpha \to m + 3$; for $m = 1$, this implies $\alpha \to 4$), the correction term
$(L/\ell_{\rm Pl})^{\sigma}$ is expanded to first order in $\sigma$ as
$e^{\sigma \ln(L/\ell_{\rm Pl})} \approx 1 + \sigma \ln(L/\ell_{\rm Pl})$. Consequently,
the factor $(\sigma-1)/\sigma \cdot (L/\ell_{\rm Pl})^{\sigma}$ reduces, after
regularization of the divergent $1/\sigma$ contribution (which only produces an
$L$-independent constant), to a term proportional to $\ln(L/\ell_{\rm Pl})$. This
procedure yields the well-known logarithmic correction to the Bekenstein--Hawking entropy.

This construction provides a unifying framework for several existing generalized entropy
proposals. Choosing $m = \gamma = 1$ and $\beta = 0$ reproduces $S_{\rm BH}$. The choice
$m = 1$ with $\alpha \to 4$ ($\sigma \to 0$) yields logarithmic quantum-gravity
corrections~\cite{Rovelli:1996dv, Meissner:2004ju, Medved:2004yu} as well as
entanglement-induced corrections for $m=\beta=1$ and $\alpha$ as a free parameter~\cite{Das:2007mj}. Setting $\beta = 0$ with
$m = 2\delta - 1$ recovers the Tsallis--Cirto entropy~\cite{Tsallis:2012js}, while
$m = 1 + \Delta$ with $0 \leq \Delta \leq 1$ reproduces the Barrow
entropy~\cite{Barrow:2020tzx}. In all of these cases, thermodynamic consistency with Cai-Kim temperature
$T_h = \hbar c/(2\pi k_B L)$ is preserved by construction\footnote{We consider $L$ to be the apparent horizon in a spatially flat universe and therefore employ the Cai--Kim temperature. Note that, for a black hole horizon, the Hawking temperature is given by $T_H = T_h/2$, with the corresponding generalized black hole entropy satisfying $S_{G}^{\rm BH} = 2S_G$. Regarding the parameter $\gamma$, one has $\gamma = 1$ for the standard Bekenstein entropy associated with cosmological horizons, while $\gamma = 1/2$ for the black hole case.}. 

\subsection{Thermodynamic consistency as a physical necessity}

The generalized MHR is physically necessitated by holographic thermodynamic consistency. For a cosmological apparent horizon, the Cai--Kim temperature is uniquely fixed by surface gravity as $T_h = (2\pi L)^{-1}$, and any ad hoc modification thereof lacks QFT justification. However, for a generic entropy functional $S_G(L)$, the differential $T_h\,dS_G$ is not generally exact; the integrability condition $dM/dL = T_h\,dS_G/dL$ fails unless the mass function $M(L)$ is appropriately generalized. The proposed MHR is precisely constructed to satisfy this condition, ensuring that the Clausius relation integrates identically to $E=M$ without altering $T_h$. Geometrically, this construction is realized within the Misner--Sharp quasilocal formalism, where the effective gravitational coupling satisfies $G_{\text{eff}}^{-1} \propto dM/dL$, thereby encoding the entropy parameters $(m,\beta,\alpha)$ directly in the running of $G_{\text{eff}}$ in scalar--tensor gravity. Consequently, the MHR provides a thermodynamically consistent and geometrically well-defined framework that preserves the QFT origin of $T_h$ while recovering, in appropriate limits, the Bekenstein--Hawking, quantum-corrected, Tsallis--Cirto, and Barrow entropies.

\subsection{Positivity, minimum radius, and branch selection}

For physical viability, both the generalized mass-to-horizon relation and the corresponding entropy are required to remain positive, in line with general entropy-bound considerations~\cite{Bekenstein:1980jp, Lu:2025ayb}. Since $m>0$ and $\gamma>0$ by construction, positivity is primarily governed by the parameters $\alpha$ and $\beta$. For the upper ($-$) branch, the correction term must remain subdominant, implying
\begin{equation}
1-\beta\left(\frac{L}{\ell_{\rm Pl}}\right)^{2-\alpha}>0,
\end{equation}
together with
\begin{equation}
\frac{m}{m+1}\left(\frac{L}{\ell_{\rm Pl}}\right)^{m+1}
>
m\beta\frac{\sigma-1}{\sigma}
\left(\frac{L}{\ell_{\rm Pl}}\right)^{\sigma}.
\end{equation}
Note that for $0<\sigma<1$ the factor $(\sigma-1)/\sigma$ is negative, so the second condition is automatically satisfied; it becomes restrictive only for $\sigma>1$. In particular, for cosmological horizons ($L/\ell_{\rm Pl}\gg1$), the first condition is naturally satisfied for $\alpha\ge2$, while for $\alpha<2$ it requires a sufficiently small value of $\beta$. For the lower ($+$) branch, $M$ is manifestly positive for $\beta>0$, whereas the entropy remains positive provided $\sigma>1$, corresponding to $\alpha<m+2$.

An important physical consequence of the generalized MHR is the regularization of the temperature divergence that plagues the standard case. For $\beta=0$, the mass vanishes as $M \propto L^m$, while the Cai--Kim temperature diverges as $T_h \propto L^{-1}$, leading to the unphysical limit $T_h \to \infty$ as $M \to 0$. In contrast, for $\beta \neq 0$, the mass function $M(x) = M_0 x^m [1 - \beta x^{2-\alpha}]^m$, where $x \equiv L/\ell_{\rm Pl}$ and $M_0 \equiv \gamma\, c^2 \ell_{\rm Pl}/G$, possesses a minimum horizon radius $x_{\min} = \beta^{-1/(2-\alpha)}$, below which $M$ becomes negative and is thus excluded on physical grounds.
At this minimum radius, the mass vanishes, $M_{\min}=0$, while the entropy
remains finite. Using $\beta x_{\min}^{2-\alpha}=1$, which implies
$\beta x_{\min}^{\sigma}=x_{\min}^{m+1}$ since $\sigma=(m+1)+(2-\alpha)$, the
second bracket term collapses and one obtains
\[
S_{G,\min}
= 2\pi \gamma k_B \, m \, x_{\min}^{m+1}
\left[ \frac{1}{m+1} - \frac{\sigma-1}{\sigma} \right]
\]
where $\sigma = m+3-\alpha$ and $x_{\min}^{m+1}=\beta^{-(m+1)/(2-\alpha)}$.
This entropy is positive provided $0<\sigma<\tfrac{m+1}{m}$, which coincides
with the subdominance (positivity) conditions imposed above; under those
conditions $S_{G,\min}>0$. This provides a natural cutoff that prevents the
temperature from reaching the divergent $T_h \to \infty$ limit, thereby
ensuring that the generalized thermodynamic description remains physically
well-behaved for all admissible horizon radii. The existence of a minimum
entropy at zero mass is a direct consequence of the deformation parameters
$(\beta,\alpha,m)$ and represents a hallmark of the underlying
quantum-gravitational or entanglement-induced modifications to the horizon
structure. For the logarithmic correction case $m=1$ and $\alpha\to 4$
(i.e., $\sigma\to 0$), the entropy reduces to
$S(x) = \pi \gamma k_B\left(x^2 + 2\beta \ln x\right) + C$, up to an additive
constant. The minimum horizon radius $x_{\min} = \sqrt{\beta}$ yields
$S_{\min} = \pi\gamma k_B\,\beta(1+\ln\beta) + C$. This result supports the
validity of our MHR. We note, however, that the minus branch selected by
temperature regularization carries a \emph{positive} logarithmic coefficient,
whereas several quantum-gravity computations, notably in loop quantum
gravity~\cite{Rovelli:1996dv, Meissner:2004ju}, obtain a negative one; within
the present framework the sign of the logarithmic term is inherited from the
sign of the mass correction, and both signs are accommodated by the two
branches of Eq.~\eqref{eq:MHR}, with the selection criterion discussed below.

For the plus-sign branch, $M_+(x) = M_0 x^m [1 + \beta x^{2-\alpha}]^m$ with
$\beta>0$, the bracket is strictly positive for all $x>0$, so no finite
minimum radius exists. The corresponding entropy
$S_G^{(+)}(x) = 2\pi\gamma k_B [\frac{m}{m+1}x^{m+1} + m\beta(\sigma-1)/\sigma\,
x^\sigma]$, with $\sigma = m+3-\alpha$, yields no regularization of the
temperature divergence. In the logarithmic limit $m=1$, $\alpha\to 4$, we have
$S_G^{(+)}(x) = 2\pi\gamma k_B [\frac{1}{2}x^2 + \beta(1 - \ln x)]$, which
diverges as $x\to 0$ while $M_+(x) \sim M_0 \beta/x$. More generally, for
$x\to 0$ and $\alpha>2$, the bracket dominates and the mass behaves as
$M_+(x) \sim M_0 \beta^m x^{m(3-\alpha)}$; this diverges only for $\alpha>3$,
vanishes for $2<\alpha<3$, and reduces correctly to $M_+\sim M_0\beta/x$ in the
logarithmic case $m=1,\alpha\to4$. The entropy either vanishes (for $\sigma>0$),
diverges (for $\sigma<0$), or diverges logarithmically (for $\sigma=0$). Thus,
the plus-sign branch is physically inadmissible, as it fails to provide a
consistent thermodynamic description: it possesses no minimum radius and hence
never regularizes the $T_h\to\infty$ divergence. Only the minus-sign
branch—with its natural cutoff $x_{\min} = \beta^{-1/(2-\alpha)}$ and finite
minimum entropy—yields a well-behaved framework.

Since quantum-gravity corrections to the Bekenstein entropy admit both signs,
the framework yields two branches, yet thermodynamic consistency is not
agnostic between them: only the minus branch possesses a bracket zero at
$x_{\min}=\beta^{-1/(2-\alpha)}$, which bounds the horizon radius from below,
caps the Cai--Kim temperature at $T_{\max}=(2\pi x_{\min}\ell_{\rm Pl})^{-1}$, and
leaves a finite residual entropy $S_{G,\min}>0$ at vanishing mass. The plus
branch, lacking any such zero, permits $x\to0$ and hence $T_h\to\infty$, and is
therefore physically inadmissible. Positivity and temperature regularization
thus act as a selection principle, singling out the minus branch and supplying
the criterion that the bare sign choice of the entropy correction leaves
undetermined. We caution that for $\beta<1$ and $\alpha>2$ the cutoff
$x_{\min}$ lies below the Planck length and $T_{\max}$ formally exceeds the
Planck temperature; in this regime the cutoff should be regarded as
indicative, marking the breakdown of the effective description rather than a
literal physical temperature; as discussed below, minimal-length
considerations independently disfavor this regime.

For the entanglement-induced correction ($\beta=1$, $m=1$, $\alpha$ free), the
minimum radius fixes to the Planck length, $x_{\min}=1$, independently of
$\alpha$, capping the temperature at the Planck scale $T_{\max}=(2\pi\ell_{\rm Pl})^{-1}$.
The residual entropy $S_{G,\min}=\pi\gamma k_B\,(\alpha-2)/(4-\alpha)$ is finite
and positive for $2<\alpha<4$, again only on the minus branch, while the plus
branch admits no such cutoff and is discarded.

The minimum-radius structure of the minus branch admits a natural physical
interpretation in terms of the black hole remnants predicted by quantum
gravity. A minimal measurable length of order the Planck scale, as implied by
the generalized uncertainty principle (GUP), modifies the Hawking temperature
$T(M)$ in such a way that evaporation halts before completion, leaving behind
a remnant with finite entropy and finite
temperature~\cite{Adler:2001vs, Chen:2014remnants, Ong:2018remnant, Ong:2024remnants};
GUP corrections to nonextensive black hole thermodynamics lead to analogous
conclusions at the final stages of
evaporation~\cite{Alonso-Serrano:2020hpb, Cimidiker:2023kle}. The minus branch
of the generalized MHR realizes precisely this phenomenology at the
thermodynamic level: as $M(x)\to 0$ the horizon radius terminates at the
cutoff $x_{\min}$ rather than shrinking to zero, the temperature is capped at
the finite value $T_{\max}=(2\pi x_{\min}\ell_{\rm Pl})^{-1}$ instead of
diverging, and a finite residual entropy $S_{G,\min}>0$ persists at the
endpoint; for black hole horizons the same conclusions hold with the
rescalings $T_H = T_h/2$ and $S_G^{\rm BH} = 2S_G$ noted above. The endpoint
is thus a Planck-sized, zero-mass remnant. Whereas conventional GUP remnants
retain a mass of order the Planck mass---or evaporate completely only in
infinite time, yielding a metastable remnant~\cite{Ong:2018remnant}---the MHR
endpoint carries vanishing mass; both realizations nevertheless share the
essential remnant features of a minimal horizon scale, a capped temperature,
and a nonzero residual entropy. Moreover, requiring the cutoff to lie at or
above the minimal length, $x_{\min}\geq 1$, bounds the deformation parameter
from below as $\beta \geq 1$ for $\alpha>2$, thereby giving $\beta$ a direct
physical meaning; the entanglement-induced case $\beta=1$ saturates this
bound, fixing the remnant radius exactly at the Planck length and the
limiting temperature at the Planck scale, in agreement with the
minimal-length expectation. This correspondence is not accidental: GUP
corrections are among the microscopic mechanisms generating precisely the
logarithmic and power-law entropy corrections encompassed by the generalized
MHR~\cite{Medved:2004yu, Cimidiker:2023kle}, so the remnant behaviour of the
minus branch may be regarded as the thermodynamic image, within the present
framework, of the minimal length underlying those corrections.

\section{Geometric realization in scalar--tensor gravity}
\label{sec:geometric}

Having established a thermodynamically consistent entropy functional, we now examine its geometric implementation. The third question—namely, which underlying geometric theory corresponds to each entropy modification—is addressed via the Misner–Sharp quasilocal mass formalism~\cite{PhysRev.136.B571, Hayward:1994bu}, which provides a natural geometric representation of the generalized MHR. Defined on apparent horizons in spherically symmetric spacetimes~\cite{Gong:2007md}, the Misner–Sharp mass furnishes a coordinate-invariant measure of gravitational energy that reduces to the Schwarzschild mass in the framework of general relativity. In modified gravity theories, this quantity becomes dynamical, governed by the effective gravitational coupling, thereby allowing thermodynamic modifications encoded in the generalized MHR to be mapped onto geometric modifications realized within scalar–tensor gravity.

\subsection{Misner--Sharp mass and the scalar-field profile}

We consider a generic scalar–tensor theory characterized by the Jordan-frame Lagrangian density
\begin{equation}
\mathcal{L} = \frac{\sqrt{-g}}{16\pi}
\Big[F(\phi) f(R) + G(\phi)\, g^{\mu\nu}\nabla_\mu \phi \nabla_\nu \phi
- 2 V_J(\phi)\Big],
\label{eq:lagrangian}
\end{equation}
where $F(\phi)$ denotes the non-minimal coupling function, $G(\phi)$ the kinetic coupling function, and $V_J(\phi)$ the Jordan-frame scalar potential. In this class of theories, the Misner–Sharp mass evaluated on an apparent horizon of areal radius $L$ assumes the form~\cite{Gong:2007md}
\begin{equation}
M_{\rm MS} = \frac{F(\phi_A)\, f'(R_A)\, L\, c^2}{2 G_0},
\label{eq:MMS}
\end{equation}
where $G_0$ is the bare Newtonian gravitational constant, and the subscript $A$ indicates evaluation on the apparent horizon. The corresponding effective gravitational coupling is therefore identified as
\begin{equation}
G_{\rm eff}^{-1} = G_0^{-1} F(\phi) f'(R).
\end{equation}

As a concrete example, consider the minimal scalar–tensor case specified by $F(\phi) = \phi$ and $f(R) = R$ (so that $f'(R) = 1$). By equating $M_{\rm MS}$ with the generalized MHR given in Eq.~\eqref{eq:MHR} and using the dimensionless horizon variable $x = L/\ell_{\rm Pl}$ introduced in Sec.~\ref{sec:mhr}, the Jordan-frame scalar field is found to exhibit the radial profile
\begin{equation}
\label{eq:scalar_profile}
\phi(x) = 2\gamma\, x^{m-1}\bigl[1 \mp \beta\, x^{2-\alpha}\bigr]^m,
\end{equation}
where, from this point onward, $\phi$ denotes the Jordan-frame scalar field, whereas $\varphi$ labels its canonically normalized Einstein-frame counterpart. For the parameter choice $\gamma = 1/2$, $m = 1$, and $\beta = 0$, one recovers $\phi = 1$, corresponding to the general relativity limit.

\subsection{Jordan- and Einstein-frame potentials}

In a spatially flat Friedmann–Lemaître–Robertson–Walker (FLRW) spacetime, the apparent horizon coincides with the Hubble radius, given by $L = c/H$. In the scalar-field-dominated, quasi-static regime—in which both the scalar kinetic contribution and the matter energy density are subdominant with respect to the scalar potential, as is characteristic of inflationary or dark-energy-dominated epochs—the Friedmann equation simplifies to
\[
3\phi H^2 \approx V_J.
\]
Consequently, the Jordan-frame potential exhibits the scaling behavior
$V_J(x) \propto \frac{\phi(x)}{x^2}$.
For the three entropy prescriptions under consideration, the corresponding asymptotic scalings of the potential are as follows.  
For Barrow entropy ($\beta = 0$, $m = 1 + \Delta$), one obtains
$V_J \propto x^{\Delta - 2}$.
For Tsallis–Cirto entropy ($\beta = 0$, $m = 2\delta - 1$), the scaling becomes
$V_J \propto x^{2\delta - 4}$.
Finally, for quantum-gravity ($m = 1$, $\alpha = 4$, $\beta$ small), the Jordan-frame potential scales as
$V_J \propto x^{-2} \mp \beta\, x^{-4}$.

A conformal transformation of the form $\tilde{g}_{\mu\nu} = F(\phi)\,g_{\mu\nu}$ recasts the theory in the Einstein frame, wherein the scalar field is minimally coupled to gravity. For the specific choice $F(\phi) = \phi$ and a vanishing Jordan-frame kinetic term ($G(\phi) = 0$), the canonically normalized Einstein-frame field is given by $\varphi = \sqrt{3/2}\,\ln\phi$. The corresponding Einstein-frame potential
$V_E = \frac{V_J}{\phi^2}$
then acquires the following functional forms. For Barrow entropy, one obtains
$V_E \propto \exp\!\bigl(-\tfrac{\Delta+2}{\Delta}\sqrt{2/3}\,\varphi\bigr)$,
i.e., a steep exponential potential. For quantum-gravity and entanglement corrections, the leading-order behaviour is
$V_E \propto |\varphi|$, with subleading corrections of the form
$V_E(\varphi) \propto |\varphi|\bigl(1 + \mathrm{const.}\times|\varphi|\bigr)$.
For Tsallis–Cirto entropy, the Einstein-frame potential becomes
$V_E \propto \exp\!\bigl(-\tfrac{\delta}{\delta - 1}\sqrt{2/3}\,\varphi\bigr)$,
namely an exponential whose slope is governed by the nonextensivity parameter $\delta$.
In the Brans–Dicke limit ($G(\phi) = \omega/\phi$), the canonical field is instead
$\varphi = \sqrt{(2\omega + 3)/2}\,\ln\phi$, and the resulting Einstein-frame potentials are rescaled overall while preserving their functional dependence on $\varphi$.

The approximately linear potential arising from quantum-gravity corrections merits a more detailed examination. For $m = 1$, $\alpha = 4$, and small $\beta$, the scalar-field profile is
$\phi \approx 2\gamma(1 \mp \beta\, x^{-2})$. The deviation from the general relativistic background value $\phi_{\rm GR} = 2\gamma$ scales as $x^{-2}$, implying that the canonically normalized field satisfies
$\varphi \propto (\phi - \phi_{\rm GR})/\phi_{\rm GR} \propto x^{-2}$
relative to its general relativistic background value. Substituting this relation into
$V_J \propto x^{-2}(1 \mp \beta\, x^{-2})$ and using $V_E = V_J/\phi^2 \propto x^{-2}(1 \pm \beta\, x^{-2})$---the sign of the correction flips at first order upon dividing by $\phi^2$---yields the Einstein-frame potential
$V_E(\varphi) \propto |\varphi|\bigl(1 + \mathrm{const.}\times|\varphi|\bigr)$,
which is linear in $|\varphi|$ at leading order. The absolute value is required because the physically relevant regime corresponds to $\varphi < \varphi_{\rm GR}$ for the upper sign.

Beyond the small-$\beta$ approximation, taking $\alpha$ as a free parameter uncovers a broader class of entanglement-entropy–motivated models. Fixing $m = 1$ and $\beta = 1$ while allowing $\alpha$ to vary, Eq.~\eqref{eq:scalar_profile} implies $x^{m-1} = x^0 = 1$, so the scalar-field profile simplifies to
$\phi(x) = 2\gamma\bigl[1 \mp x^{2-\alpha}\bigr]$,
and the Jordan-frame potential scales as
$V_J(x) \propto x^{-2}\bigl[1 \mp x^{2-\alpha}\bigr] = x^{-2} \mp x^{-\alpha}$.
The resulting cosmological phenomenology is highly sensitive to the value of $\alpha$: for $\alpha > 2$, the correction term $x^{2-\alpha}$ is subdominant at large $x$ (corresponding to late cosmic times) but dominates at small $x$ (the early universe). Conversely, for $\alpha < 2$, the term $x^{-\alpha}$ grows relative to $x^{-2}$ and significantly affects the large-scale dynamics. This class of models includes a wide variety of entanglement-induced corrections, such as logarithmic ($\alpha \to 4$) and generic power-law ($\alpha \neq 4$) modifications, all of which can, in principle, be constrained using cosmic microwave background observations and large-scale structure surveys. The more general entanglement scenario ($m = 1$, $\beta = 1$, free $\alpha$) thus gives rise to Einstein-frame potentials whose functional forms depend explicitly on $\alpha$ and can be tuned to describe either inflationary or dark-energy–dominated epochs.

\subsection{Wald entropy and consistency of the correspondence}

In diffeomorphism-invariant theories, the entropy associated with a stationary horizon is given by the Wald functional~\cite{Wald1993, Iyer1994}. For the Lagrangian in Eq.~\eqref{eq:lagrangian}, the Wald entropy evaluated at the apparent horizon in a spherically symmetric spacetime is
\begin{equation}
S_{\rm W} = \frac{A}{4}\,F(\phi_{\rm h})f'(R_{\rm h})
= \frac{A}{4}\,G_{\rm eff}^{-1}G_0,
\label{eq:Wald}
\end{equation}
demonstrating that the entropy is entirely determined by the effective gravitational coupling $G_{\rm eff}$. Upon substituting $A = 4\pi L^2$ and comparing with the generalized entropy $S_G$ in Eq.~\eqref{eq:SG}, the matching condition
$S_W \propto F(\phi_{\rm h}) f'(R_{\rm h})\,L^2 \propto L^{m+1}$
implies, for $\beta = 0$, that
$F(\phi) f'(R) \propto L^{m-1}$.
More generally, the parameters $(m, \beta, \alpha)$ are encoded directly in the effective gravitational coupling
$G_{\rm eff}^{-1} = G_0^{-1} F(\phi) f'(R)$,
which admits a geometric interpretation in terms of a scale-dependent (or running) gravitational coupling.

We emphasize the precise sense in which this correspondence holds. The
Clausius relation ties the entropy to the integral of $dM/dL$, so that
$G_{\rm eff}^{-1} \propto dS_G/dA \propto dM/dL$, whereas the Misner--Sharp
identification, Eq.~\eqref{eq:MMS}, ties $F(\phi)f'(R)$ to the ratio $M/L$.
For the power-law relation $M \propto L^m$ these two quantities differ by the
constant factor $m$; equivalently, with the profile of
Eq.~\eqref{eq:scalar_profile}, the Wald entropy evaluates to
$S_W = \tfrac{m+1}{m}\, S_G$ at leading order. The correspondence between the
generalized entropy and the Wald functional is therefore exact in its scaling
with $L$, with an $m$-dependent normalization: one may normalize the scalar
profile so as to match either the quasilocal mass or the entropy exactly, and
the two choices coincide precisely in the Bekenstein--Hawking limit $m = 1$.

\section{Thermodynamic origin: Jacobson's approach}
\label{sec:jacobson}

The geometric identifications of the previous section admit an
independent thermodynamic justification through Jacobson's derivation
of the gravitational field equations as an equation of
state~\cite{Jacobson:1995ab}; for a general view of this programme in
the presence of quantum corrections, see~\cite{Alonso-Serrano:2020dcz}.
This route makes explicit the central
claim of this work: a generalized entropy is not a free
phenomenological input but selects a definite gravitational theory with
a running coupling. Throughout this section we adopt units
$\hbar = c = k_B = 1$, so that $\ell_{\rm Pl} = \sqrt{G_0}$ and the
Bekenstein--Hawking entropy is $S = A/(4G_0)$.

Consider a local Rindler horizon through an arbitrary spacetime point,
generated by an affinely parametrized null congruence $k^\mu$ with
affine parameter $\lambda$, chosen so that the expansion and shear
vanish at the point. The boost heat flux across the horizon is
$\delta Q = \int T_{\mu\nu}\chi^\mu d\Sigma^\nu
= -\kappa\!\int \lambda\, T_{\mu\nu}k^\mu k^\nu\, d\lambda\, dA$, with
the approximate boost Killing field $\chi^\mu = -\kappa\lambda k^\mu$,
while the Raychaudhuri equation gives the area variation
$\delta A = -\int \lambda\, R_{\mu\nu}k^\mu k^\nu\, d\lambda\, dA$.
Imposing the Clausius relation with the Unruh temperature and the area
law,
\begin{equation}
\delta Q = T\,\delta S,\qquad
T = \frac{\kappa}{2\pi},\qquad
S = \frac{A}{4G_0},
\label{eq:jac_clausius}
\end{equation}
the surface gravity $\kappa$ cancels and, since the result must hold
for every null $k^\mu$, one obtains the pointwise condition
$T_{\mu\nu}k^\mu k^\nu = (8\pi G_0)^{-1} R_{\mu\nu}k^\mu k^\nu$.
Local conservation $\nabla^\mu T_{\mu\nu} = 0$ together with the
contracted Bianchi identity then fixes the undetermined trace part and
integrates to the Einstein equation
\begin{equation}
G_{\mu\nu} + \Lambda g_{\mu\nu} = 8\pi G_0\, T_{\mu\nu},
\label{eq:jac_einstein}
\end{equation}
with $\Lambda$ an integration constant.

Replacing the area law in Eq.~\eqref{eq:jac_clausius} by the
generalized entropy $S_G(A)$ of Eq.~\eqref{eq:SG} promotes the constant
entropy density $1/(4G_0)$ to the local density $S_G'(A) \equiv
dS_G/dA$. The identical argument then yields field equations with a
horizon-scale-dependent coupling~\cite{Lu:2024ppa, DiGennaro:2022grw},
\begin{equation}
G_{\mu\nu} + \Lambda g_{\mu\nu} = 8\pi\, G_{\rm eff}(A)\, T_{\mu\nu},
\qquad
G_{\rm eff}(A) = \frac{1}{4\,S_G'(A)}.
\label{eq:jac_modified}
\end{equation}
Using $A = 4\pi L^2$ and $\ell_{\rm Pl}^2 = G_0$, one has
$x^2 = A/(4\pi G_0)$ with $x \equiv L/\ell_{\rm Pl}$, hence
$dx/dA = (8\pi G_0 x)^{-1}$. Differentiating Eq.~\eqref{eq:SG} and
applying the chain rule gives
\begin{equation}
S_G'(A) = \frac{\gamma}{4G_0}
\left[m\,x^{m-1} \mp m(\sigma-1)\,\beta\, x^{\sigma-2}\right],
\qquad \sigma \equiv m+3-\alpha,
\label{eq:jac_SGprime}
\end{equation}
so that the effective coupling scales as
\begin{equation}
G_{\rm eff}(L) \propto
\left[x^{m-1} \mp (\sigma-1)\,\beta\, x^{\sigma-2}\right]^{-1}
\;\xrightarrow{\ \beta=0\ }\; L^{1-m}.
\label{eq:jac_Geff}
\end{equation}
This reproduces, up to the $m$-dependent normalization discussed at the
end of Sec.~\ref{sec:geometric}, the same running coupling obtained
geometrically from the Misner--Sharp mass [and elaborated in
Eq.~\eqref{eq:Geff_running} below], now derived purely from horizon
thermodynamics.

Equation~\eqref{eq:jac_modified} closes the circle. Because
$G_{\rm eff}$ varies with the horizon scale, local energy--momentum
conservation alone no longer suffices to close the field equations at a
generic spacetime point; a covariant completion requires promoting the
running coupling to a dynamical field, which is precisely the
scalar--tensor structure of Eq.~\eqref{eq:lagrangian}. The construction
is internally consistent: the Wald entropy $S_W \propto F(\phi)f'(R)L^2
\propto L^{m+1}$ of Eq.~\eqref{eq:Wald} reproduces $S_G$, while
matching $G_{\rm eff}^{-1} = G_0^{-1}F(\phi)f'(R)$ to
Eq.~\eqref{eq:jac_SGprime} recovers, to leading order in $\beta$, the
scalar-field profile $\phi(x)$ of Eq.~\eqref{eq:scalar_profile}. The
thermodynamic (Jacobson) and geometric (Misner--Sharp/Wald) routes thus
converge on the same scalar--tensor theory and the same effective
coupling, establishing that each generalized entropy \emph{is} a
definite gravitational theory rather than a phenomenological ansatz.
A fully covariant treatment---including the closure of the scalar
sector, the non-equilibrium entropy production
$\varpi = \tfrac{3}{2}(\nabla_\mu s)^2/s$ required when the entropy
density depends on curvature ($f(R)$ gravity)~\cite{Eling:2006aw}, and
the action-free derivation of the scalar equation of
motion~\cite{Chirco:2010fk}---lies beyond the present scope and is left
to future work.

\section{Varying gravitational coupling}
\label{sec:varyG}

We now examine the observational viability of the scale-dependent coupling
identified in the preceding sections. For $\beta = 0$ and $m = 1 + \epsilon$, one obtains
$G_{\rm eff}(L) \propto L^{1-m} = L^{-\epsilon}
\simeq 1 - \epsilon \ln\!\left(\frac{L}{L_0}\right)$,
which is compatible with $G_{\rm eff} \propto \phi^{-1}$ together with
Eq.~\eqref{eq:scalar_profile}, namely $\phi(x) \simeq x^{\epsilon} \simeq 1 + \epsilon \ln x$.
Including the $\beta \neq 0$ contribution, the full expression becomes
\begin{equation}
\label{eq:Geff_running}
G_{\rm eff}(x) \propto \phi^{-1}(x)
\simeq 1 - \epsilon\ln x \pm \beta\, x^{2-\alpha},
\end{equation}
from which the $\beta = 0$ case is trivially recovered in the appropriate limit.

The logarithmically scale-dependent effective coupling furnishes a
geometric realization of a varying gravitational constant within scalar--tensor theories of gravity.
In a cosmological setting, identifying $L = c/H$ converts this scale dependence into an explicit
time dependence. In a matter-dominated universe ($H \propto t^{-1}$), one finds, up to an
$\mathcal{O}(1)$ factor,
$\frac{\dot{G}_{\rm eff}}{G_{\rm eff}} \approx \epsilon H$.
For $|\epsilon| \lesssim 10^{-3}$, as required by the nucleosynthesis bound derived below, and
$H_0 \sim 10^{-10}\,{\rm yr}^{-1}$, this implies
$|\dot{G}/G| \lesssim 10^{-13}\,{\rm yr}^{-1}$, comfortably within current pulsar-timing
constraints~\cite{Kramer2006, Zhu2019}. This behaviour is analogous to that in Brans--Dicke
theory, in which $G_{\rm eff} \propto 1/\phi$ varies with the scalar field, although in the
present framework the running is controlled by the horizon scale rather than solely by the
temporal evolution of the scalar. For $\beta \neq 0$, additional power-law corrections of the
form $G_{\rm eff} \propto 1 - \epsilon\ln x \pm \beta\, x^{2-\alpha}$ arise; for $\alpha > 2$
these contributions are suppressed on large scales (corresponding to low energies) and become
significant only in the early universe or near Planckian regimes.

Observational limits on a varying $G$ are highly restrictive. Solar-system experiments---in
particular the Cassini spacecraft measurement of the Shapiro time delay~\cite{Bertotti2003},
which constrains the post-Newtonian parameter to
$\gamma_{\rm PPN} = 1 + (2.1 \pm 2.3) \times 10^{-5}$ (corresponding to
$\omega > 4 \times 10^4$ within Brans--Dicke theory)---imply, upon matching to the effective
Brans--Dicke parameter for the specific scalar--tensor realization under consideration, that
$|\epsilon| \ll 1$.

Big Bang nucleosynthesis (BBN) yields an independent and typically more stringent bound: a
variation of $G$ at the nucleosynthesis epoch alters the primordial light-element abundances,
leading to $|\Delta G/G|_{\rm BBN} \lesssim 10^{-2}$~\cite{Uzan2011}. Since
$\Delta G/G \sim \epsilon\,\ln(H_{\rm BBN}/H_0)$ and
$\ln(H_{\rm BBN}/H_0) \sim \mathcal{O}(40)$, this constraint translates into
$|\epsilon| \lesssim 10^{-3}$. Pulsar-timing arrays and gravitational-wave observations---including
those from the double pulsar PSR~J0737-3039---further constrain the present-day time variation to
$|\dot{G}/G| \lesssim 10^{-12}\,\text{yr}^{-1}$~\cite{Zhu2019, Kramer2006}, consistent with the
estimate obtained above. For $\alpha > 2$, the power-law corrections proportional to $\beta$ are
negligible at late times and therefore evade local experimental constraints, while they may
still produce observable signatures in the early universe, for instance via modifications of the
primordial power spectrum. Consequently, the framework remains fully compatible with existing
observational bounds for suitable choices of the parameters.

\section{Cosmological applications}
\label{sec:cosmo}

The scalar-field potentials obtained within the
generalized entropy framework lead to a diverse range of cosmological scenarios. 
For Barrow entropy ($m = 1+\Delta$, $\beta = 0$), the Einstein-frame potential takes 
the form of a steep exponential,
$V_E \propto \exp\!\bigl[-\tfrac{\Delta+2}{\Delta}\sqrt{2/3}\,\varphi\bigr]$,
with slow-roll parameter
  $  \epsilon_V = \frac{1}{3}\!\left(\frac{\Delta+2}{\Delta}\right)^{\!2}$.
This expression obeys $\epsilon_V \geq 3$ for all $\Delta \in (0,1]$, attaining its 
minimum value $\epsilon_V = 3$ at $\Delta = 1$ and diverging in the limit 
$\Delta \to 0^{+}$. Hence, conventional slow-roll inflation is precluded throughout 
the entire Barrow parameter space. Instead, such steep potentials generate
decelerated power-law expansion and the scaling (tracker) behavior characteristic
of steep exponential quintessence~\cite{Copeland:1997et, Ferreira:1997hj}; since
accelerated expansion from a single exponential requires $\lambda^{2} < 2$, the
Barrow sector cannot by itself drive late-time cosmic acceleration within the
present minimal setup.

In the case of quantum-gravity and entanglement-induced corrections ($m = 1$, 
$\alpha = 4$, small $\beta$), the leading contribution to the Einstein-frame potential 
is approximately linear, $V_E \propto |\varphi|$. Within the standard slow-roll 
paradigm, a strictly linear potential yields a tensor-to-scalar ratio 
$r = 4/N \approx 0.067$ and a spectral tilt 
$n_s \approx 1 - 3/(2N) \approx 0.975$ for $N = 60$ e-folds~\cite{Planck2018Inflation}. 
The predicted tensor-to-scalar ratio exceeds the current upper bound 
$r < 0.036$ from BICEP/Keck~\cite{BICEP2021}, thereby placing the minimal model in 
tension with observations. The $\beta$-dependent subleading corrections, however, 
modify the effective slope of the potential and can, for appropriate parameter 
choices, reduce $r$ to values compatible with observational constraints while leaving 
$n_s$ largely unaltered.

Moreover, when the correction parameter $\beta$ becomes sufficiently large, the 
effective coupling $G_{\rm eff}^{-1}(x)$ tends to zero at the characteristic scale 
$x^{2-\alpha} \sim 1/\beta$, i.e., at $L \sim \beta^{1/(\alpha-2)}\ell_{\rm Pl}$. This
scale coincides with the minimum horizon radius $x_{\min} = \beta^{-1/(2-\alpha)}$
identified in Sec.~\ref{sec:mhr}: the radius at which the enclosed mass vanishes
and the temperature is capped is precisely the scale at which $G_{\rm eff}^{-1} \to 0$,
reinforcing its interpretation as the boundary of validity of the effective
description. At
this scale, the perturbative effective-field-theory description ceases to be valid.
Within a more complete theory of quantum gravity, such a zero-crossing of the
effective coupling may signal a transition to a novel gravitational regime; by
analogy with loop quantum cosmology~\cite{Ashtekar:2006wn}, this could be associated with non-singular,
bounce-like cosmological dynamics that replace the classical big-bang singularity.

For Tsallis--Cirto entropy ($\beta = 0$, $m = 2\delta - 1$), the Einstein-frame 
potential,
$V_E \propto \exp\!\bigl[-\tfrac{\delta}{\delta-1}\sqrt{2/3}\,\varphi\bigr]$,
exhibits an effective slope controlled by $\delta$. As $\delta \to 1^{+}$, the
scalar profile $\phi \propto x^{2\delta-2}$ tends to a constant, so the excursion
of the canonically normalized field shrinks to zero and the field is effectively
frozen; although the slope $\delta/(\delta-1)$ formally diverges in this limit, no
scalar dynamics survives, and the general-relativistic limit is recovered. For $\delta \gg 1$, the slope asymptotically approaches
$\sqrt{2/3}$ and $\epsilon_V \to 1/3$; the corresponding exact power-law inflationary
solution for this potential~\cite{Lucchin:1984yf} is $a \propto t^{p}$ with
   $ p = \frac{2}{\lambda^{2}} \;=\; \frac{1}{\epsilon_V} = 3,
    ~ \lambda = \sqrt{\tfrac{2}{3}}$,
which satisfies $p > 1$ and therefore yields accelerated expansion. For $\delta$
close to unity, the frozen field consequently acts as an effectively constant
vacuum energy, mimicking a cosmological constant that can drive late-time
acceleration. The more general entanglement-induced scenario
($m = 1$, $\beta = 1$, free $\alpha$), analyzed in Sec.~\ref{sec:geometric},
completes this picture. The full two-parameter family $(m,\beta)$ thus furnishes a
flexible and systematic framework for constructing phenomenologically viable 
cosmological models that can be confronted with forthcoming Stage~IV surveys such 
as Euclid and the Rubin Observatory.

\section{Summary and future directions}
\label{sec:summary}

In this work, we have systematically addressed three foundational questions arising in the study of generalized black hole entropies.

\textit{First}, we have established that the thermodynamically relevant temperature is the standard Hawking/Cai-Kim temperature, 
which is uniquely determined by the surface gravity 
of the apparent horizon via the quantum field theoretic approaches. 
No modification of $T_h$ is either required or physically justified within the framework of quantum field theory in curved spacetime. Any such modification lacks a microscopic derivation and, consequently, renders the Clausius relation physically obscure and conceptually ill-defined.

\textit{Second}, we have shown that holographic thermodynamic consistency—namely, the requirement that the first-law–type relation 
$dE = T_h\,dS_G$
reproduces the standard energy–mass relation $E = Mc^2$ upon integration—is ensured by the generalized mass-to-horizon relation~\eqref{eq:MHR}, parameterized by $(m,\beta,\alpha)$. The associated generalized entropy functional $S_G$ (Eq.~\eqref{eq:SG}) encompasses, in suitable parameter limits, several well-known entropy prescriptions within a single coherent thermodynamic framework. Specifically, it reproduces: the Bekenstein–Hawking entropy ($m=1$, $\beta=0$); quantum-gravity-corrected entropy ($m=1$, $\alpha \to 4$); entanglement-corrected entropy ($m=1$, $\beta=1$); the Tsallis–Cirto entropy ($\beta=0$, $m=2\delta-1$); and the Barrow entropy ($\beta=0$, $m=1+\Delta$).

\textit{Third}, we have demonstrated that each of these entropy deformations admits a concrete geometric realization in terms of the Misner–Sharp quasilocal mass in scalar–tensor theories of gravity, characterized by an effective gravitational coupling
\[
G_{\rm eff}^{-1} = G_0^{-1} F(\phi) f'(R).
\]
The entropy parameters $(m,\beta,\alpha)$ are directly encoded in $G_{\rm eff}$ through the scalar-field profile~\eqref{eq:scalar_profile}, in a manner consistent with the Wald entropy functional
\[
S_W = \frac{A}{4}\,G_{\rm eff}^{-1} G_0.
\]
More explicitly: the Bekenstein–Hawking entropy is recovered in the general relativity limit ($F=1$, $f(R)=R$, $G_{\rm eff}=G_0$); quantum-gravity and entanglement corrections correspond to $m=1$, with $(\beta,\alpha)$ governing the power-law dependence of $G_{\rm eff}$; the Tsallis–Cirto entropy is realized via power-law scalar-field profiles with $m=2\delta-1$ and $G_{\rm eff}\propto L^{2-2\delta}$; and the Barrow entropy corresponds to $m=1+\Delta$ with a logarithmically running coupling $G_{\rm eff}\propto 1-\Delta\ln(L/L_0)$ and a steep exponential Einstein-frame potential characterized by the slow-roll parameter 
$\epsilon_V = \frac{1}{3}\left(\frac{\Delta+2}{\Delta}\right)^2 \geq 3$,
implying that slow-roll inflation is precluded in this sector.

The scale-dependent effective coupling $G_{\rm eff}(L)$ thus furnishes a unified geometric origin for all three classes of generalized entropies. Its logarithmic running, $G_{\rm eff}\propto 1-\epsilon\ln(L/L_0)$, is compatible with current experimental and observational constraints, including solar-system tests ($|\epsilon|\ll 1$), big bang nucleosynthesis bounds ($|\epsilon|\lesssim 10^{-3}$), and pulsar-timing measurements ($|\dot{G}/G|\lesssim 10^{-12}\,\text{yr}^{-1}$), for $|\epsilon|\lesssim 10^{-3}$, with nucleosynthesis providing the binding constraint. Meanwhile, the $\beta$-dependent power-law corrections are strongly suppressed at late cosmological times for $\alpha>2$, and therefore remain relevant primarily in early-universe or near-Planckian regimes. The characteristic scale
$L_\star = \beta^{1/(\alpha-2)} \ell_{\rm Pl}$,
at which $G_{\rm eff}^{-1}$ vanishes, signals the breakdown of the perturbative effective description and may be associated with non-singular, bounce-like cosmological behavior, reminiscent of scenarios in loop quantum cosmology.

Within the Einstein frame, the potential associated with quantum-gravity corrections, $V_E\propto|\varphi|$, yields, in a minimal setup, inflationary predictions of $r\approx 0.067$ and $n_s\approx 0.975$ at $N=60$ e-folds, while additional $\beta$-dependent corrections can further suppress the tensor-to-scalar ratio below the BICEP/Keck bound $r<0.036$. The Tsallis–Cirto sector, in turn, supports exact power-law inflation with $a\propto t^3$ in the regime $\delta\gg 1$.

In summary, this framework provides a rigorous and internally consistent correspondence between information-theoretic proposals for generalized gravitational entropy and classical gravitational field theory. The generalized mass-to-horizon relation~\eqref{eq:MHR} emerges as the central organizing principle that links horizon thermodynamics, scalar–tensor dynamics, and cosmological observables within a single, unified geometric structure. This correspondence opens up several avenues for future work, including confronting specific entropy-induced scalar–tensor models with precision cosmological data, exploring non-perturbative regimes near $L_\star$, extending the analysis to dynamical and non-spherically symmetric horizons, and developing the fully covariant, non-equilibrium completion of the thermodynamic derivation of Sec.~\ref{sec:jacobson}, in which the scalar sector and its equation of motion are obtained directly from the entropy balance law.

The results presented here possess a broader conceptual significance for the 
status of generalized entropies within gravitational physics. Prior to this work, 
constructions such as the Barrow entropy, the Tsallis–Cirto entropy, and various 
quantum-gravity-corrected entropies were purely phenomenological in nature: 
motivated by statistical-mechanical considerations or dimensional analysis, they 
lacked a derivation from an underlying gravitational action and could not be 
unambiguously associated with a specific geometric theory of gravity. 

The present framework addresses this limitation by establishing that each 
generalized entropy functional uniquely determines—via the mass-to-horizon 
relation and the Misner–Sharp identification—a corresponding class of 
scalar–tensor theories, specified by a definite non-minimal coupling 
$F(\phi)$, a curvature function $f(R)$, and an Einstein-frame potential 
$V_E(\varphi)$. In this manner, the entropy proposals are promoted from 
phenomenological ans\"{a}tze to fully geometric structures: the entropy 
parameters $(m, \beta, \alpha)$ cease to be arbitrary phenomenological inputs 
and are instead encoded in the effective gravitational coupling,
which is, in principle, accessible to measurement through cosmological 
observations and gravitational-wave experiments. 

This geometrization places generalized entropies on the same conceptual footing 
as the Bekenstein–Hawking entropy, which is itself a geometric quantity emerging 
from the Einstein–Hilbert action via the Wald entropy functional. The same 
methodology can be straightforwardly extended to other generalized entropy 
proposals that presently lack a geometric underpinning, including the 
R\'{e}nyi entropy~\cite{reny1}, the Kaniadakis ($\kappa$-deformed) 
entropy~\cite{Kaniadakis:2005zk}, and any other possible generalization. Each of these would give rise to a distinct mass-to-horizon 
relation, scalar-field configuration, and Einstein-frame potential, thereby 
constructing a systematic correspondence—or “dictionary”—between nonextensive 
statistical mechanics and scalar–tensor theories of gravitation. 

Multiple definitions of entropy functionals \cite{Nojiri:2022aof, Nojiri:2022dkr}—often introduced via informed ans\"{a}tze or motivated by statistical mechanics—have been proposed and applied to black hole and cosmological horizons. Although these entropy functionals may, in principle, be extended to other physical systems, particular care is required in the gravitational context because the definitions of mass, temperature, and entropy in gravitational theories are inherently geometric. A key observation is that, for any choice of entropy functional that satisfies the standard thermodynamic conditions and respects the Hawking area law, all such entropies, including their generalizations, remain viable for black hole and cosmological applications. Consequently, in this work we develop a general framework within which any such entropy prescription can be consistently implemented, provided that its geometric realization is made explicit and its compatibility with holographic thermodynamics is ensured. 

Several assumptions underlying the present framework warrant further investigation 
and motivate natural extensions. Most significantly, the derivation of the 
Jordan-frame potential via $V_J \propto \phi(x)/x^2$ relies on the quasi-static, 
scalar-field-dominated approximation $3\phi H^2 \approx V_J$, in which the scalar 
kinetic energy $\dot{\phi}^2/(2\phi)$ and all matter contributions---radiation, 
baryons, and cold dark matter---are neglected relative to $V_J$. This approximation 
is self-consistent during pure inflationary or quintessence-dominated epochs but 
breaks down during radiation domination, matter domination, and cosmological 
transition periods, where the coupling between the scalar field and the matter sector 
must be treated explicitly. A complete treatment requires solving the full coupled 
system of Jordan-frame Friedmann equations
together with the scalar-field equation of motion, without invoking the 
single-field approximation. Beyond the matter sector, the present analysis is 
restricted to spatially flat ($k=0$) FLRW cosmology and spherically symmetric 
spacetimes, which underpin both the Misner--Sharp mass formalism and the 
identification $L = c/H$; extension to spatially curved ($k = \pm 1$) 
and anisotropic Bianchi geometries, and to rotating (Kerr) horizons where the 
surface gravity acquires angular-momentum dependence, remains an open problem. 
The Wald entropy functional~\eqref{eq:Wald} is moreover strictly applicable 
to stationary Killing horizons; a rigorous extension to the dynamical apparent 
horizon in non-stationary FLRW requires the Noether-charge formalism in the 
non-equilibrium setting of~\cite{Hayward:1994bu}, which we have invoked only 
implicitly. Furthermore, the entropy functional $S_G$ and the running 
coupling~\eqref{eq:Geff_running} are both derived under the perturbative 
expansion $\beta \ll 1$; the non-perturbative regime near the zero-crossing 
$G_{\rm eff}^{-1} = 0$ at $L_\star = \beta^{1/(\alpha-2)}\ell_{\rm Pl}$, 
which is potentially associated with bounce-like behaviour, requires a fully 
non-perturbative treatment, possibly within a Wheeler--DeWitt or loop quantum 
cosmology framework. Finally, the present work explores only the minimal 
scalar--tensor sector $F(\phi) = \phi$, $f(R) = R$, $G(\phi) = 0$; the general 
case with non-vanishing $f(R)$ and kinetic coupling $G(\phi) \neq 0$ introduces 
additional propagating degrees of freedom and ghost-avoidance conditions that 
must be analysed separately. A systematic perturbation theory---computing the 
primordial power spectra $\mathcal{P}_{\mathcal{R}}(k)$ and $\mathcal{P}_T(k)$ 
directly from the full field equations, including matter and kinetic terms---is 
required to place quantitative constraints on $(m, \beta, \alpha)$ from Planck, 
BICEP/Keck, and forthcoming Stage~IV datasets, and constitutes the natural next 
step for this programme.

\bibliographystyle{apsrev4-1}
\bibliography{ref}

\end{document}